\documentclass[11pt]{article}
\usepackage[a4paper,margin=1in,top=25mm]{geometry}
\usepackage{amsmath,amssymb,amsthm}
\usepackage{graphicx}
\usepackage{hyperref}
\usepackage{authblk}

\title{Open Nanoacoustic Resonators Based on SrTiO$_3$/YBa$_2$Cu$_3$O$_{7-x}$ Superlattices}
\author[1]{S. Sandeep}
\author[1]{O. Colmegna}
\author[2]{S. Carreira}
\author[2]{L. M. Vicente-Arche}
\author[3]{L. B. Steren}
\author[2]{J. Briatico}
\author[1]{N. D. Lanzillotti-Kimura\thanks{Corresponding author: daniel.kimura@cnrs.fr}}

\affil[1]{Université Paris-Saclay, C.N.R.S., Centre de Nanosciences et de Nanotechnologies (C2N),  10 Boulevard Thomas Gobert, 91120 Palaiseau, France}
\affil[2]{Laboratoire Albert Fert, CNRS, Thales, Université Paris-Saclay, 91767 Palaiseau, France}
\affil[3]{Instituto de Nanociencia y Nanotecnología, CNEA-CONICET, Centro Atómico Constituyentes, San Martín, Argentina}

\begin{document}

\maketitle

\begin{abstract}
We report the design and experimental demonstration of an open nanophononic cavity based on a hybrid oxide superlattice composed of SrTiO$_3$ (STO) and YBa$_2$Cu$_3$O$_{7-x}$ (YBCO), combined with a metallic Ni transducer for coherent phonon generation. The STO/YBCO periodic stack acts as an acoustic distributed Bragg reflector supporting confined longitudinal acoustic phonons in the sub-THz regime, while the Ni layer enables efficient ultrafast optical excitation and detection by time-domain Brillouin scattering. Transient reflectivity measurements reveal confined acoustic dynamics and a well-defined cavity resonance, in agreement with transfer-matrix calculations of acoustic reflectivity and mode profiles. These results demonstrate phonon confinement in multifunctional oxide heterostructures and establish complex oxide superlattices as a platform for hybrid nano-acoustic resonators and ultrafast phonon control of correlated electronic phases.
\end{abstract}

\section{Introduction} \label{introduction}

The development of advanced nanofabrication and optical characterization techniques, particularly time-domain Brillouin scattering (TDBS), has enabled the investigation of acoustic phonons in the gigahertz-to-terahertz regime, opening opportunities for applications in thermal management, ultrafast optoelectronics, and quantum technologies ~\cite{priyaPerspectivesHighfrequencyNanomechanics2023}. Among the different nanophononic platforms, superlattices (SLs) are especially attractive because their periodic structure modifies the phonon dispersion relation through Brillouin-zone folding and the formation of phononic minigaps ~\cite{colvardObservationFoldedAcoustic1980,bartelsCoherentZoneFolded1999}. As a result, acoustic SLs can function as frequency-selective filters~\cite{narayanamurtiSelectiveTransmission1979,ndlkNanowaveDevicesTerahertz2006}, for controlling phonon transport~\cite{yaremkevichProtectedLongDistanceGuiding2021,xiangwaveguide2024a}, as reflective elements in Fabry–Pérot resonators ~\cite{huynhSubterahertzPhononDynamics2006}, or as components designed for the simultaneous control of hypersound and electromagnetic waves~\cite{fainsteinStrongOptMech2013}. Moreover, Brillouin-zone folding makes high-frequency phonons optically addressable.

While most studies have focused on semiconductor systems such as GaAs/AlAs ~\cite{NdlkTowardsGHzTHzcavity2015,priyaPerspectivesHighfrequencyNanomechanics2023} and SiGe ~\cite{ezzahriCoherentPhononsInSiSiGesuperlattices2007}, these materials have benefited from mature fabrication techniques, making them reliable platforms for both fundamental and applied research. At the same time, alternative platforms including porous materials ~\cite{abdalaMesoporousThinFilms2020,cardozodeoliveiraProbingGigahertzCoherent2023,boggianoOpticalReadoutMechanical2023,cardozoEnvResp2026, lomonosovNanoscaleNoncontactSubsurface2012,sandeepLowK2022}, colloidal and nanoparticle assemblies ~\cite{chengNanospher2006,akimovOpal3d2008,ayouchAssemblyDisord2012,graczykowski2020,abighanemProgressLU2024,morontaC2N2025}, hybrid soft--hard systems~\cite{schneiderHybrid2013,alonsoPhoxonic2015}, correlated and phase-change materials~\cite{ mogunovVO2PhaseT2020,guGeTe2021}, and magnetic heterostructures~\cite{ scherbakovMagnetization2010, thevenard2010, thevenardSawMagSwit2016,stupakiewiczUltrafastPhononicSwitching2021,vlasovModernProblemsUltrafast2022} have broadened the accessible functionalities in nanophononics.

Oxide superlattices are particularly appealing because they enable control of physical properties through the interplay of lattice, charge, and other degrees of freedom, leading to emergent phenomena not present in bulk materials ~\cite{rameshCreatingEmergentPhenomena2019}. Previous TDBS studies on oxide heterostructures have reported coherent acoustic phonon dynamics, confined cavity modes, and coupled acoustic resonances in systems such as YBCO/LCMO, BTO/STO/Ni, and STO/BTO/SRO~\cite{liCoherentAcousticPhonons2016,lanzillotti-kimuraEnhancementInhibitionCoherent2010,bruchhausenAcousticConfinementPhenomena2018}. More recently, BFO/LFO multiferroic superlattices have exhibited additional folded and terahertz phonon modes arising from structural ordering beyond the chemical periodicity, highlighting the potential of oxide superlattices for tuning coherent THz phonons and related excitations such as magnons~\cite{guSuperordersTerahertzAcoustic2024}.

In this work, we investigate coherent acoustic phonon generation and confinement in STO/YBCO superlattices and hybrid Ni-capped structures using TDBS. We demonstrate the realization of an open acoustic cavity defined by the superlattice mirror and the Ni/air interface. Confined modes in the gigahertz range are experimentally observed and reproduced by transfer-matrix calculations. These results provide insight into phonon confinement and tunability in complex oxide heterostructures, with prospects for studying changes in acoustic properties across phase transitions in YBCO.

\section{Materials and Methods} \label{methods}

\begin{figure}[htbp]
    \centering\includegraphics[width=7cm]{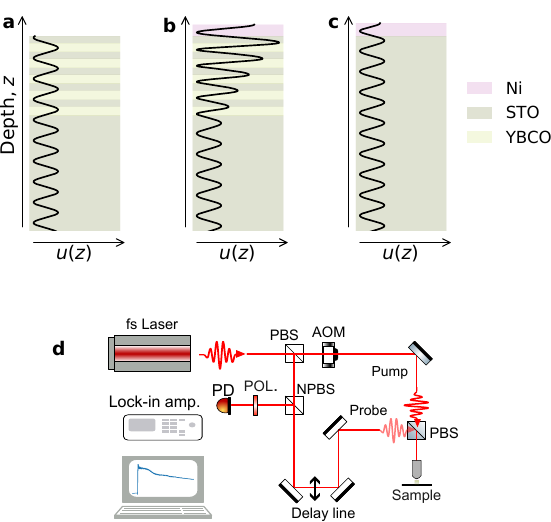}
    \caption{Schematic of the investigated STO/YBCO heterostructures and the TDBS experimental setup. (a) STO/YBCO superlattice grown on an STO substrate. (b) Hybrid Ni/STO/YBCO structure forming an open nanoacoustic cavity through the combination of the Ni/air interface and the STO/YBCO acoustic mirror. (c) Control Ni/STO sample without the superlattice. The black curves represent the acoustic displacement field $u(z)$ at $\sim 110$~GHz, evidencing confinement of the cavity mode only in the hybrid structure. (d) TDBS pump--probe setup used for generation and detection of coherent acoustic phonons.}
    \label{fig1}
\end{figure}
\subsection{Sample growth and characterization}
Three types of samples were examined to study the acoustic properties of oxide heterostructures  (Figure~\ref{fig1}(a--c)). All multilayers were grown on single-crystal SrTiO$_3$ (001) substrates using pulsed laser deposition (PLD). The first sample consists of a SrTiO$_3$/YBa$_2$Cu$_3$O$_7$ superlattice (SL) comprising five periods of alternating layers with nominal thicknesses of 11~nm for SrTiO$_3$ and 14~nm for YBa$_2$Cu$_3$O$_7$. This structure, hereafter referred to as the SL sample, serves as a reference sample to describe the acoustic phonon generation and detection in SL using  TDBS. The second sample is an open-cavity structure (OC) based on a similar superlattice structure, consisting of five periods of SrTiO$_3$ (11~nm) and YBa$_2$Cu$_3$O$_7$ (14~nm). A metallic Ni film was deposited ex-situ on top of the superlattice by DC sputtering, with nominal thicknesses of 20, 25, 30, and 35~nm across different regions of the sample. This configuration enables tuning of the acoustic phonon confinement through boundary conditions defined by the superlattice and the free surface (air interface), resulting in a thickness-dependent resonance behavior. The third sample is a control structure consisting of a single Ni film (20~nm thick) deposited directly on a SrTiO$_3$ substrate. This sample serves as a control for disentangling the role of the superlattice in phonon confinement.
The layer periodicity and interface quality of all samples were characterized by X-ray reflectivity (XRR), while the structural quality was analyzed by x-ray diffraction (XRD).

\subsection{Time-resolved optical measurements}

Transient reflectivity measurements were carried out using a degenerate pump--probe setup driven by a Ti:Sa laser (Coherent Chameleon Ultra II) with a pulse duration of approximately 140 fs and a repetition rate of 80 MHz (Figure~\ref{fig1}(d)). The central wavelength was set around 840 nm. Cross-polarized pump and probe were focused on the sample to a spot size of around 2 microns. The reflected probe beam was detected using a photodetector, while the reflected pump was filtered based  on polarization. An acousto-optic modulator was used to modulate the pump at 800 kHz as a reference for the lock-in detection, improving the sensibility to reflectivity changes following the initial excitation.

\section{Results and Discussion} \label{results}
\subsection{Acoustic response of the superlattice}
\begin{figure}[!t]
    \centering\includegraphics[width=7cm]{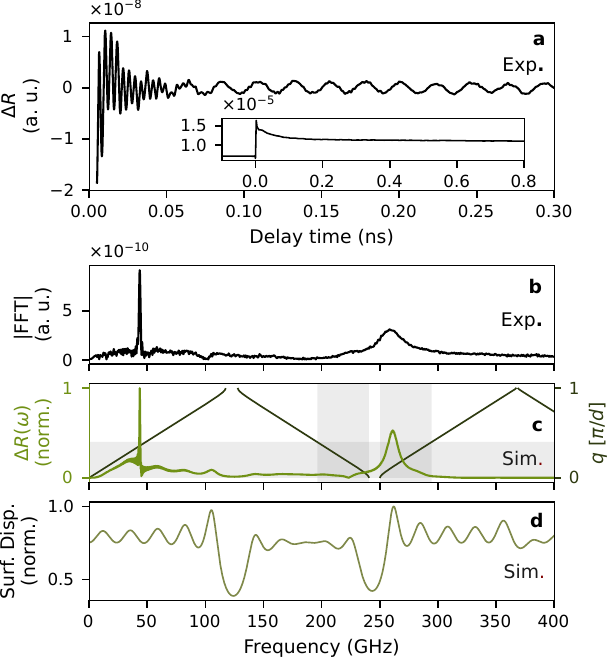}
    \caption{Acoustic response of the STO/YBCO SL measured by TDBS and simulations. (a) Background-subtracted transient reflectivity signal showing coherent acoustic oscillations generated in the SL together with Brillouin oscillations (inset: raw transient reflectivity signal). (b) Fourier transform of the experimental signal. (c) Simulated acoustic detection spectrum $\Delta R(\omega)$ obtained from the transfer-matrix model, together with the calculated acoustic dispersion relation of an infinite STO/YBCO SL (black line). The shaded region indicates the uncertainty in wavevector arising from the finite thickness of the SL. (d) Calculated surface displacement profile of the structure.}
    \label{SL_results}
\end{figure}

Figure ~\ref{SL_results}(a) shows the acoustic component of the transient reflectivity signal observed in the SL sample. The acoustic signal arises from a strain pulse generated within the structure, induced by the femtosecond pump pulse. Strain generation occurs in the YBCO layers via a thermoelastic mechanism, as previously reported ~\cite{kashiwadaSituMonitoringGrowth2006}. The probe pulse detects the acoustic response through changes in reflectivity. The acoustic signal is extracted by subtracting the slowly varying background that appears in the raw transient reflectivity signal (inset of Figure ~\ref{SL_results}(a)) after the sharp change in reflectivity at zero time delay. This background is associated with the electronic and thermal contributions induced by the pump excitation. The resulting signal consists of decaying high-frequency oscillations accompanied by lower-frequency oscillations. The Fourier spectrum of the acoustic signal is shown in Figure ~\ref{SL_results}(b). The dominant low-frequency oscillation is observed at $43.6\,\mathrm{GHz}$, which corresponds to the Brillouin frequency in the STO layer, given by $f_B = \frac{2 n v}{\lambda_{\mathrm{laser}}}$. Using a refractive index of 2.34 for STO, the extracted sound velocity is $\sim 7.8\,\mathrm{nm/ps}$, consistent with previously reported values ~\cite{brivioObservatioSto2011,sandeepStoBtoEvalOptAcou2021}. These oscillations arise from propagating acoustic waves in both the multilayer structure and the substrate. Due to the relatively small thickness of the SL structure, the dominant contribution at later times—after the decay of high-frequency oscillations—originates from longitudinal acoustic waves propagating in the substrate. The high-frequency oscillations observed in the initial time window are attributed to superlattice acoustic modes.

To analyze the origin of the observed high frequency modes, we compare the experimental results with simulations based on the transfer matrix method for a one-dimensional multilayer structure ~\cite{ndlkCohGenAcoustic2007,pascual-winterSpectralResponses2012}. The spectral features observed in the transient reflectivity signal originate from both the photoelastic interaction associated with the generated strain and the displacement of interfaces at the surface and within the multilayer structure ~\cite{matsuda2002}. The simulated detection spectrum corresponding to the photoelastic contribution is presented in Figure ~\ref{SL_results}(c) and shows good agreement with the experimental observations, reproducing both the dominant low-frequency peak ($\sim 43.6\,\mathrm{GHz}$) and the high-frequency feature ($\sim 260\,\mathrm{GHz}$). The high-frequency peak in the experiment exhibits a shift of approximately $2\%$ relative to the simulation. This discrepancy may arise from uncertainties in layer thicknesses or material parameters, particularly the sound velocity in thin films.

To further interpret these modes, the dispersion relation of an infinite SL is shown in Figure ~\ref{SL_results}(c). The dominant SL mode at $\sim 260\,\mathrm{GHz}$ is located near the wave vector $q = 0$ in the folded Brillouin zone, within an uncertainty $\Delta q = \frac{2\pi}{L}$ imposed by the finite thickness of the SL ~\cite{pascual-winterSpectralResponses2012}. Among the zone-center modes, the dominant contribution is determined by the symmetry of the acoustic displacement and strain fields ~\cite{pascual-winterSpectralResponses2012}. In addition, the finite size of the structure gives rise to nonvanishing contributions from other band-edge modes ~\cite{ezzahriCoherentPhononsInSiSiGesuperlattices2007}, which are observed in both the simulations and the experimental results. Figure ~\ref{SL_results}(d) shows the surface displacement response, whose dominant peaks coincide with those observed in the photoelastic detection spectrum. These results demonstrate the capability of the SL structure to generate high-frequency phonons. The additional broadening observed in the experimental peaks can be attributed to attenuation of high-frequency components and the finite time window used in the Fourier transform, effects that are not included in the present simulations.

\subsection{Hybrid nanocavity structure and thickness-dependent response}
\begin{figure}[ht]
    \centering\includegraphics[width=7cm]{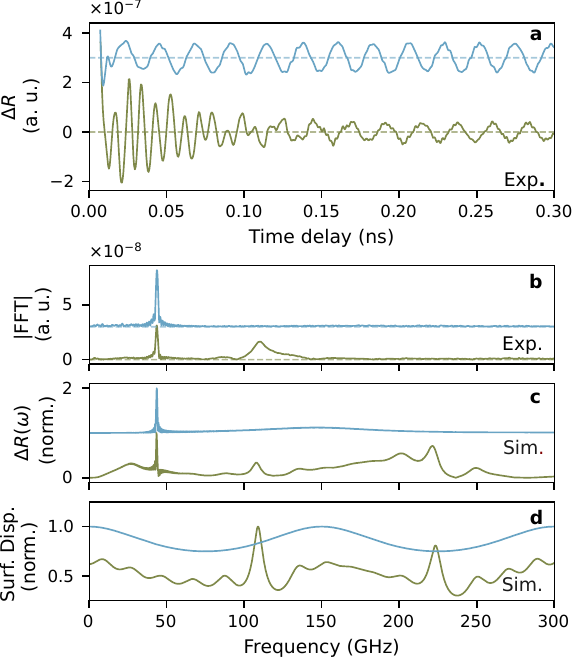}
    \caption{ Acoustic response of the hybrid open-cavity and control samples measured by TDBS and simulations. (a) Background-subtracted transient reflectivity signals for the hybrid Ni/STO/YBCO open-cavity structure (green) and the control Ni/STO sample (blue). (b) Fourier transforms of the experimental signals, showing the additional confined mode around $\sim 110$~GHz in the hybrid structure, absent in the control sample. (c) Simulated acoustic detection spectra $\Delta R(\omega)$ obtained from the transfer-matrix model. (d) Calculated normalized surface displacement spectra, evidencing acoustic confinement in the hybrid open-cavity structure.}
    \label{fig3}
\end{figure}

A hybrid nanocavity structure was realized by depositing a metallic Ni layer on top of the STO/YBCO SL. The introduction of the Ni layer modifies the acoustic response of the SL by forming a cavity near the sample surface. The large acoustic impedance contrast at the Ni/air interface imposes an effective stress-free boundary condition, resulting in strong reflection of acoustic waves. In this configuration, the SL acts as a distributed Bragg reflector and, together with the Ni/air interface, defines an acoustic cavity. Figure~\ref{fig1}(b) schematically illustrates the open cavity structure together with the calculated acoustic displacement profile of a mode confined near the surface at $\sim$110~GHz. The effective cavity corresponds to a $\lambda/2$ resonator formed by the Ni layer and the uppermost layers of the SL.

Time-resolved pump--probe measurements were performed on this hybrid structure, consisting of the STO/YBCO SL capped with a 20~nm Ni layer, under the same experimental conditions used for the reference SL sample. In this case, acoustic phonon generation occurs mainly in the metallic Ni layer due to its strong optical absorption. The transient reflectivity signal obtained from the hybrid structure is shown in Figure ~\ref{fig3}(a) and compared with that of a control sample consisting of a 20~nm Ni film directly deposited on a STO substrate (without the SL). The time-domain response clearly distinguishes the hybrid cavity structure from the control sample. The corresponding Fourier spectra are presented in Figure ~\ref{fig3}(b). The control sample exhibits a single dominant peak at 43.6~GHz, corresponding to Brillouin oscillations associated with acoustic waves propagating in the STO substrate. In contrast, the hybrid structure shows an additional resonance at $\sim$110~GHz. This frequency lies within the first stop band of the SL mirror, i.e., at the Brillouin zone-edge gap, indicating the formation of a confined cavity mode. The simulated reflectivity spectra associated with the photoelastic response and the calculated surface displacement are presented in Figure~\ref{fig3}(c) and (d), respectively, for both the hybrid and control structures. The simulations reproduce the experimentally observed resonance within the first stop band and reveal a pronounced enhancement of the surface displacement in the hybrid structure, consistent with acoustic confinement in the cavity. In contrast, the control sample does not exhibit such confinement. The experimentally observed resonance exhibits a frequency deviation of approximately 10\% relative to the simulated value. This larger deviation, compared to the $\sim$2\% shift observed for the SL sample, may be attributed to uncertainties in the STO/YBCO multilayer parameters used in the simulations, particularly the layer thicknesses and interface quality. The larger discrepancy observed for the OC sample is consistent with the increased interface roughness revealed by the XRR characterization (see supplemental document). A second resonance is expected near $\sim$220~GHz, corresponding to the second stop band, but is not observed experimentally. Its absence may originate from interface imperfections, as well as from enhanced attenuation of higher-frequency phonons in the SL layers.

The observation can be understood in terms of acoustic impedance matching. Since Ni has an acoustic impedance relatively close to that of STO, phonons generated in the Ni film efficiently transmit into the substrate, leading to weak confinement. In contrast, the SL acts as a reflective mirror, enabling phonon confinement within the hybrid cavity. Similar enhancement and inhibition of coherent phonons in Ni films have been previously reported in STO/BTO systems and on STO substrates ~\cite{lanzillotti-kimuraEnhancementInhibitionCoherent2010}, supporting the conclusion that the STO/YBCO SL acts as an acoustic mirror enabling phonon confinement in the hybrid cavity.

\begin{figure}[ht]
    \centering\includegraphics[width=7cm]{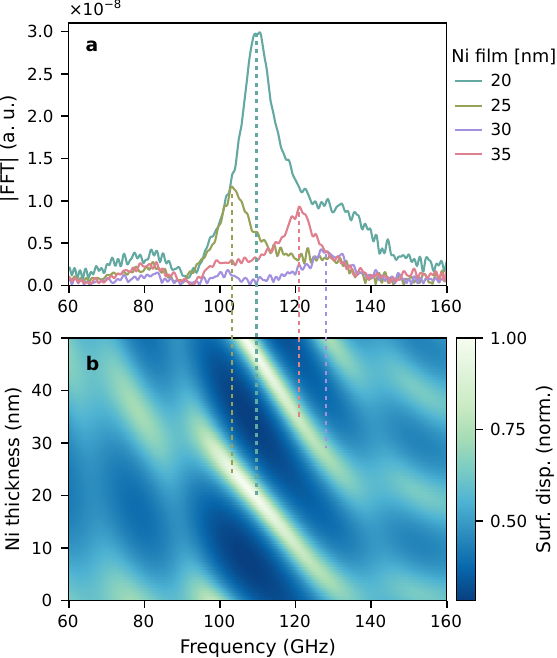}
    \caption{Thickness-dependent acoustic response of the hybrid Ni/STO/YBCO open-cavity structure. (a) Experimental Fourier spectra obtained from TDBS measurements for different Ni thicknesses, showing the systematic shift of the confined cavity mode with increasing Ni thickness. The labels indicate nominal Ni thicknesses in nm. (b) Simulated normalized surface displacement map as a function of frequency and Ni thickness}
    \label{fig4}
\end{figure}

Further evidence of cavity behavior is provided in Figure~\ref{fig4}, where hybrid structures with different Ni thicknesses were investigated. The Fourier spectra of the experimental signals are compared with simulated surface displacements for Ni thicknesses ranging from 20 to 35~nm. The results show a systematic shift of the cavity resonance frequency with increasing Ni thickness, demonstrating tunability of the confined mode. This behavior reflects the dependence of the cavity resonance on the effective cavity length and maps the stop gap at the Brillouin zone edge.

\section{Conclusion} \label{conclusion}

We have demonstrated coherent phonon generation and confinement in STO/YBCO oxide superlattices and in hybrid structures incorporating a metallic Ni layer. In the superlattice, high-frequency folded acoustic modes in the gigahertz range are observed, in agreement with transfer-matrix simulations, confirming the capability of oxide superlattices to sustain and control hypersound excitations. Beyond the structural periodicity, the system effectively behaves as a nanoacoustic SL with a well-defined modulation of the elastic properties. When combined with a Ni transducer, the system forms an open acoustic nanocavity in which the superlattice acts as a distributed Bragg reflector. This configuration gives rise to a confined mode within the first stop band in the gigahertz range, absent in the control sample without the superlattice. The resonance frequency shows a clear dependence on the Ni thickness, demonstrating tunability through the effective cavity length.

These results highlight the potential of complex oxide heterostructures as platforms for phononic engineering. In particular, the dual role of YBCO as both an acoustic transducer and a material that can sustain high-frequency acoustic phonons, together with the optical transparency of STO, opens routes toward tailored cavity designs. For instance, fully oxide-based Fabry–Pérot acoustic cavities could be realized using transparent superlattice mirrors (e.g., STO/BTO), with YBCO acting as an embedded phonon generator. Such structures would enable selective phonon generation and confinement ~\cite{pascualwinterSelectiveOpticalGeneration2007}, and provide a promising pathway to probe ultrafast dynamics and phase transitions in correlated oxide systems through high-frequency acoustic cavity modes.

\section {acknowledgement}
The authors acknowledge support from the C.N.R.S. International Research Project Phenomenas. S.S., O. C., and N.D.L.-K. acknowledge funding from European Research Council Consolidator Grant No.101045089 (T-Recs). 

\bibliographystyle{unsrt}
\bibliography{references}

\section*{Supplemental document: Structural and morphological analysis}

\setcounter{figure}{0}
\renewcommand{\thefigure}{S\arabic{figure}}
\begin{figure}[htbp]
    \centering\includegraphics[width=8cm]{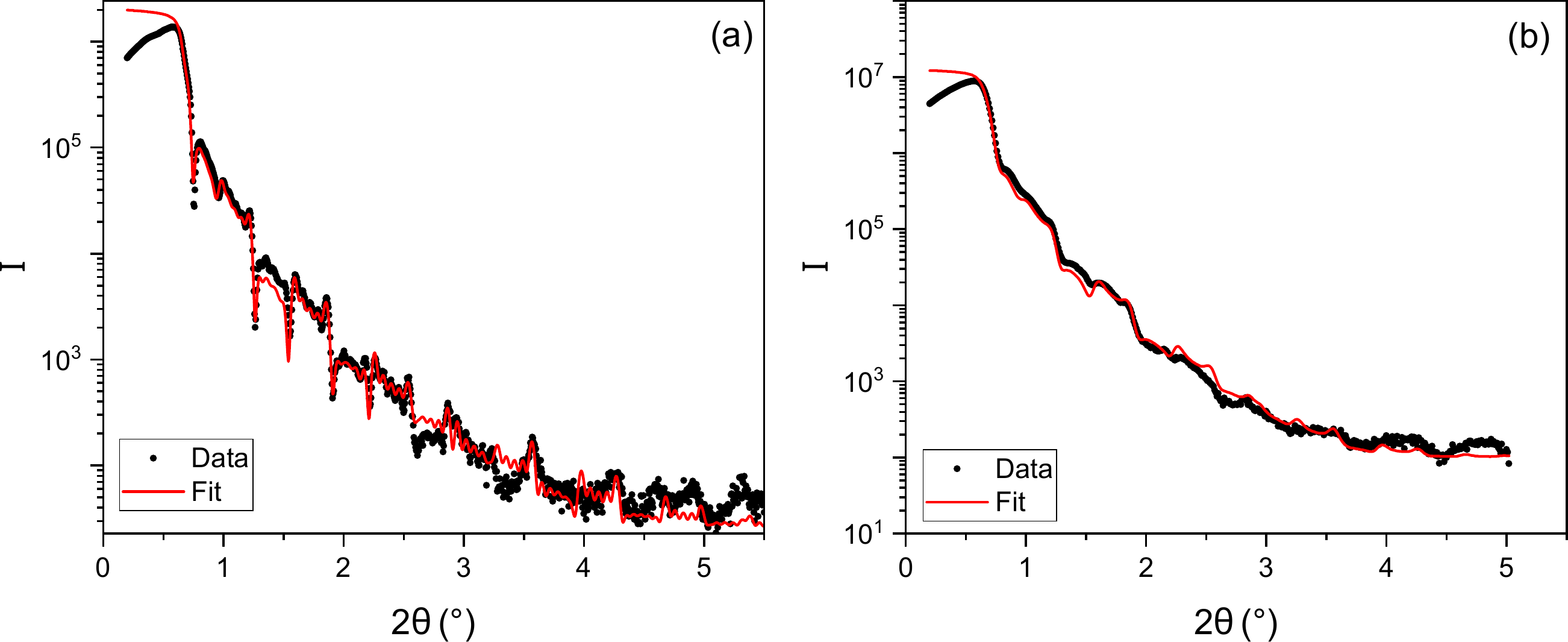}
    \caption{XRR of the (a) SL and (b) OC. The fits were obtained with GenX ~\cite{Genx3}, using the nominal thicknesses as input parameters.}
    \label{figS1}
\end{figure}

\begin{figure}[htbp]
    \centering\includegraphics[width=8cm]{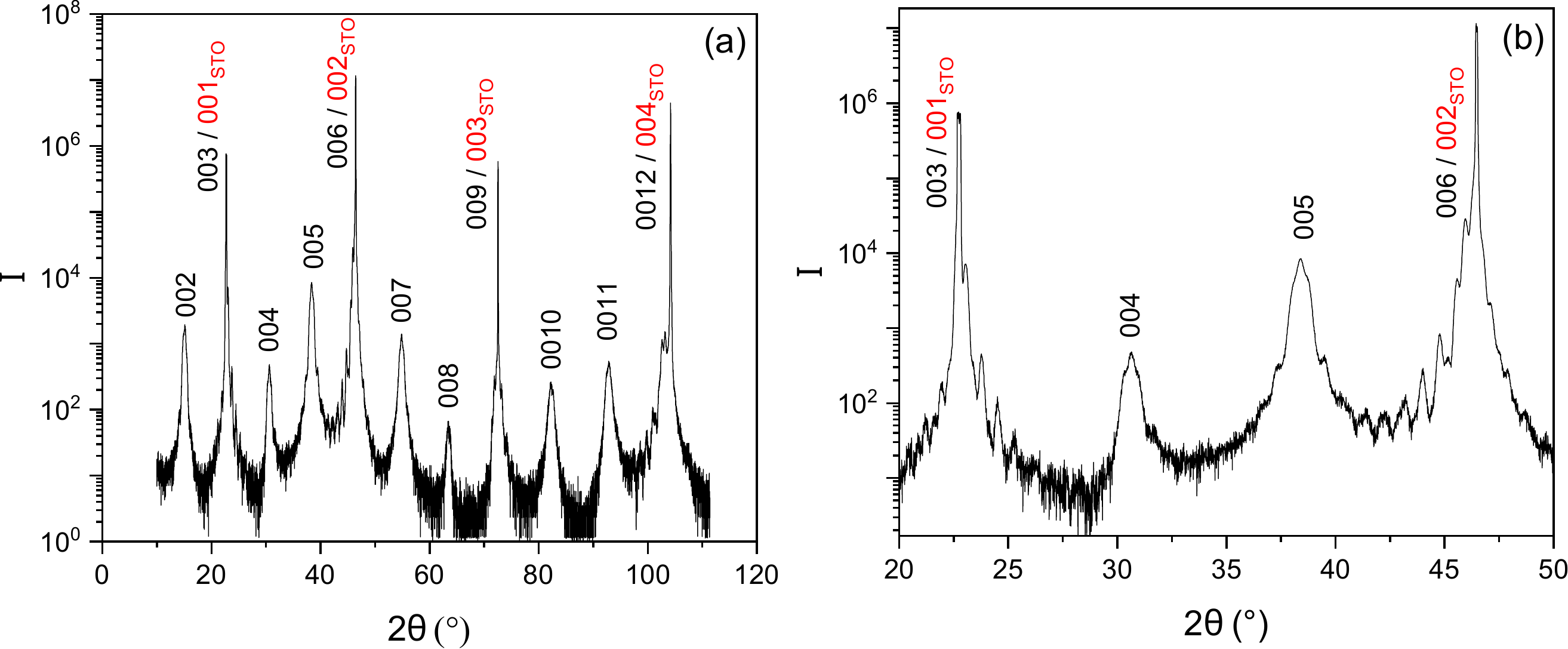}
    \caption{(a) XRD of the SL, where the (00n) peaks corresponding to the c-axis YBCO are indicated. The peaks of the STO layers are superposed to the peak of the substrates and therefore distinguishable with this technique. In (b) we show the XRD around the (001) and (002) reflections of the STO, where we can observe the Laue oscillations around the peaks.}
    \label{figS2}
\end{figure}
The interface quality and layer periodicity are key parameters for detecting the resonant properties of acoustic cavities. Moreover, a reliable simulation of the acoustic detection spectra requires well-defined thickness values and crystalline oriented structures with homogeneous stoichiometries. To assess these properties, we performed x-ray diffraction (XRD) and x-ray reflectometry (XRR). The XRR results in Figure~\ref{figS1} show well-defined oscillations consistent with superlattice structures. The data were fitted with GenX ~\cite{Genx3}, using the nominal thicknesses and the densities of the bulk materials as initial parameters. The calculated thicknesses found from the fits are similar for both samples and they are in agreement with the expected values. By comparing the data of both samples, we note a higher amplitude of the oscillations for the SL than for the OC sample. This difference can be explained by a higher roughness of the interfaces for the OC samples than for the SL and may explain the absence of the second stop band expected from the simulations near  \textasciitilde{}$\sim 220~\mathrm{GHz}$. The XRD scans in Figure~\ref{figS2} show only the $(00n)$ reflections of the YBCO, which confirms the $c$-axis orientation of the structure, and the Laue oscillations observed around the substrate'peaks confirm the high-quality interfaces.

\end{document}